\documentclass[preprint]{article}
\pdfoutput=1 
\usepackage[latin1]{inputenc}
\usepackage{color}
\usepackage[pdftex]{graphicx}
\usepackage{amsfonts}
\usepackage{subfigure}
\usepackage{amssymb}
\usepackage{amsmath}
\usepackage{latexsym}
\usepackage[left=2.5cm,top=2cm,right=2.0cm,bottom=2cm]{geometry}
\usepackage{longtable}

\begin{document}

\title{ {\huge{\bf Wavelets Applied to the Detection of Point Sources of UHECRs}}}

\author{ R. A. BATISTA$^{1\dagger}$, E. KEMP$^{1}$, R. M. de ALMEIDA$^{2}$ {\it and} B. DANIEL$^{1}$\\ 
\vspace{0.5cm}\\
$^1$Instituto de F{\'{i}}sica `` Gleb Wataghin"\\ 
Universidade Estadual de Campinas \\
13083-859, Campinas-SP, Brazil \\
\vspace{0.8cm}\\
$^2$Escola de Engenharia Industrial Metal{\'u}rgica de Volta Redonda\\
Universidade Federal Fluminense\\
27255-125, Volta Redonda-RJ, Brazil\\
\vspace{0.5cm}\\
$^\dagger$\texttt{rab@ifi.unicamp.br}\\
\vspace{0.2cm}}

\date{}

\maketitle

\vspace{0.5cm}
\begin{abstract}
    In this work we analyze the effect of smoothing maps containing arrival directions of cosmic rays with a gaussian kernel and kernels of the mexican hat wavelet family of order 1, 2 and 3. The analysis is performed by calculating the amplification of the signal-to-noise ratio for several anisotropy patterns (noise) and different number of events coming from a simulated source (signal) for an ideal detector capable of observing the full sky with equal probability. We extend this analysis for a virtual detector located within the array of detectors of the Pierre Auger Observatory, considering an acceptance law.

\end{abstract}

\vspace{1cm}

\section{Introduction}

\par The origin, chemical composition and mechanisms of acceleration  of the ultra-high energy cosmic rays (UHECRs) still remain a mistery half a century after their discovery\cite{linsley}, and are subject of great interest by the scientific community due to their intrinsic relation with particle physics, cosmology and astrophysics. The flux of cosmic rays with energies above 1 EeV ($10^{18}$ eV) is approximately 1 particle per square kilometer per year and 1 particle per square kilometer per century for energies above 60 EeV\cite{agnupdate}. 

\par To achieve a reasonable statistic of events with energies above 1 EeV, it was built in the province of Mendoza, Argentina, the Pierre Auger Observatory, the result of an international effort of 17 countries. This observatory is a pioneer in the application of the hybrid technique of simultaneous fluorescence and surface detection. The first consists on the measurement of the fluorescence light resulting from the de-excitation of the nitrogen molecules which interacted with particles from the extensive air shower (EAS) generated by the primary particle. The surface detection consists on the detection of particles from the EAS by an array of 1,600 water-Cherenkov tanks composing a triangular grid with 1.5 km spacing from each other, covering an area of approximately 3,000 km$^2$.

 \par The identification of possible astrophysical sources and the investigation of the magnetic fields which permeates the universe are studied by analyzing the arrival directions of cosmic rays. The correlation of these directions with the large scale distribution of matter in the universe, such as the galactic and supergalactic planes are considered large scale anisotropies. A small scale anisotropy is characterized by the association between arrival directions of cosmic rays and point sources, such as stars, distant galaxies and other kind of objects which are far enough to be considered point sources.

 \par Since the sources of UHECRs are still unknown, it is important to search for them. Amongst the candidates to sources of UHECRs are the active galactic nuclei, gamma-ray bursts and magnetars\cite{hillas}. Here we analyze the performance of several convolution kernels on searching for anisotropies possibly correlating to some astrophysical sources of UHECRs.

 \section{Wavelets}
 
 \par Wavelets are mathematical functions belonging to the $\mathbf{L}^2$ space that satisfy some requirements. The first one is the admissibility condition, 
 \begin{equation}
 	\int \frac{{|\psi(\omega)|^2}}{|\omega|}d\omega<\infty,
 \end{equation}
where $\psi(t)$ is the wavelet. This condition guarantees the reconstruction of a signal without loss of information. Also, the wavelet shall satisfy the condition of zero norm, i. e.,
\begin{equation}
	\int {\psi(t) dt} = 0.
\end{equation} 
This last condition means that the average value of the wavelet in time domain shall be zero at zero frequency\cite{addison}.

The continuous wavelet transform (CWT) may be formally written as:
\begin{equation}
 \Phi(s,\tau) = \int f(t)\Psi^{*}_{s,\tau}(t)dt,
\end{equation}
where $s$ ($s>0$, $s$ $\in$ $\mathbf{R}$) and $\tau$ ($\tau$ $\in$ $\mathbf{R}$) are, respectively, the scale (dilation) and translation parameters. So, the CWT decomposes a function $f(t)$ in a basis of wavelet $\Psi_{s,\tau}(t)$. The inverse transform is given by:
\begin{equation}
	f(t) = \int \int \Phi(s,\tau) \Psi_{s,\tau}(t) d\tau ds.
\end{equation}

\par These wavelets $\Psi_{s,\tau}(t)$ are obtained from a so-called {\it mother-wavelet}, by changing the dilation and translation parameters as follows:
\begin{equation}
	\Psi_{s,\tau}(t) = \frac{1}{\sqrt{s}}\Psi\left( \frac{t-\tau}{s} \right).
\end{equation}

\par The mexican hat wavelet family (MHWF) and its extension on the sphere have been widely used aiming the detection of point sources in maps of cosmic microwave background (CMB)\cite{cayon,vielva01,vielva03}, due to the amplification of the signal-to-noise ratio when going from the real space to wavelet space.

\par The MHWF is obtained by successive applications of the laplacian operator to the two-dimensional gaussian. A generic member of this family is:
\begin{equation}
	\Psi_n(\vec{x}) = \frac{(-1)^n}{2^n n!} \nabla^{2n} \phi(\vec{x}),
\end{equation}
where $\phi$ is the two-dimensional gaussian ($\phi(\vec{x}) = \frac{1}{2\pi} e^{-\vec{x}^2/2\sigma}$) and the laplacian operator is applied $n$ times.

\section{Celestial Maps}

\par When studying anisotropies, a powerful tool are the celestial maps, which are segmented pixelizations of the celestial sphere. The events map is the celestial map which represents, in two dimensions, the arrival directions of cosmic ray events in the celestial sphere.

 \par Due to limitations of the detector itself, it is impossible to determine the exact arrival direction of an event. Each event detected is convolved with a probability distribution related to the angular resolution of the detector, that is, for each event there is an associated point spreading function (PSF). Therefore, it is extremely useful to convolve the celestial maps with functions associated to the PSF of the detector, aiming to maximize the signal-to-noise ratio. Mathematically, this process of convolution (filtering\footnote{In this paper we make no distinction between the processes of convolution, smoothing and filtering.}) may be written as
 \begin{equation}
 	M_f (\vec{r_0}) = \alpha \int M(\vec{r}) \Phi(\vec{r},\vec{r_0})d\Omega,
 \end{equation}
  where $\alpha$ is a normalization constant, $M(\vec{r})$ is the number of cosmic ray events in the direction $\vec{r}$, $\Phi(\vec{r},\vec{r_0})$ is the kernel\footnote{We also make no distinction between the terms kernel and filter.} of the transformation and $\vec{r_0}$ is the position vector representing each point in which the integral is evaluated. In the discrete case this process is:
  \begin{equation}
  	M_f(k) = \frac{\sum_j M(j) \Phi(\vec{r_k},\vec{r_j})}{\sum_j \Phi(\vec{r_k},\vec{r_j})},
  \end{equation}
  where $M(j)$ is the number of cosmic rays associated to the pixel of index $j$ in the direction $\vec{r_j}$.
 
 \section{Analysis Procedure}
 
 \par In the present work we have tested the capability of the MHWF to detect point sources embedded in different backgrounds. The simulated source was simulated in the position $(l,b)$=$(320^o,30^o)$ (galactic coordinates) in the sky, with a dispersion of $2^o$. We have considered three possible scenarios for the source, varying the amplitude of the source with respect to the background, and the number of events simulated in its direction. In the first case the source is $10\%$ stronger than the background and $100,000$ events were simulated. In the second case, we have simulated $5,000$ events and the source has an amplitude of $1\%$ with respect to the background. In the third case there are only $50$ events coming from the direction of the source, which has an amplitude of $0.1\%$ with respect to the background.
 
 \par For the background we have simulated $500,000$ events according to several anisotropy patterns. These anisotropy patterns were divided in two classes: with and without the acceptance of the detector. In the case without acceptance, the distribution of events in the sky is uniform, modulated only by the anisotropy pattern imposed by the simulation. In the case with acceptance, we considered a detector in a place with latitude $35^o$ $28'$ $00''$ $S$ and longitude $65^o$  $18'$ $41''$ $W$ (approximately the coordinates of the Pierre Auger Observatory). We have assumed that the event are detected according to a zenith angle distribution that follows $sin\theta cos\theta$, where $0^o\leq \theta \leq60^o$ is the zenith angle\cite{kalcheriess}.
 
 \par
\begin{figure}[h!]
	\centering
 	\hspace{-1.7cm}
 	\includegraphics[scale=0.245]{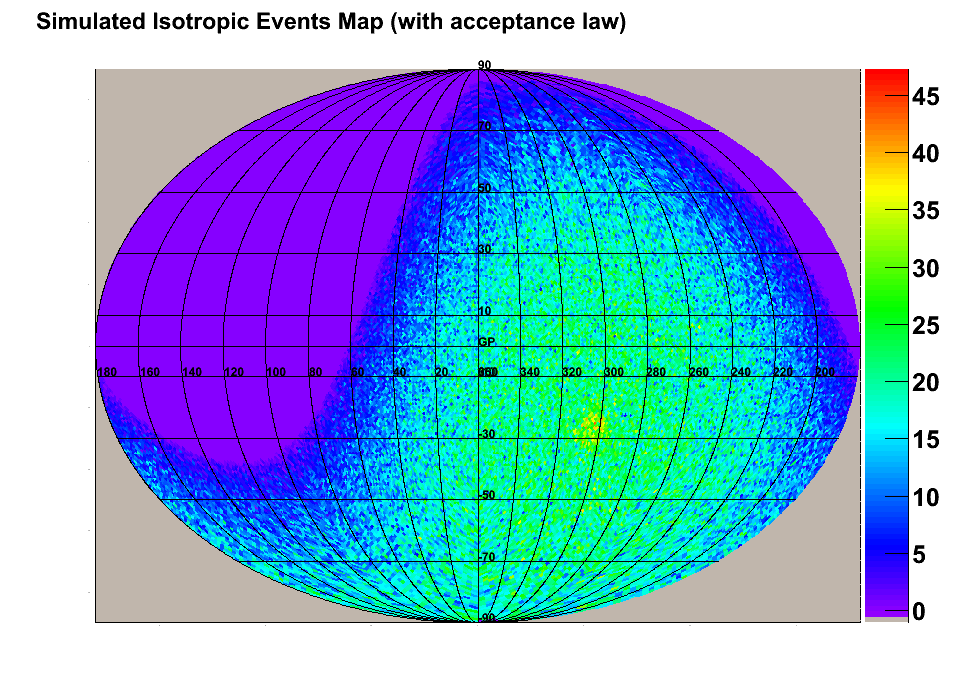} 
 	\includegraphics[scale=0.245]{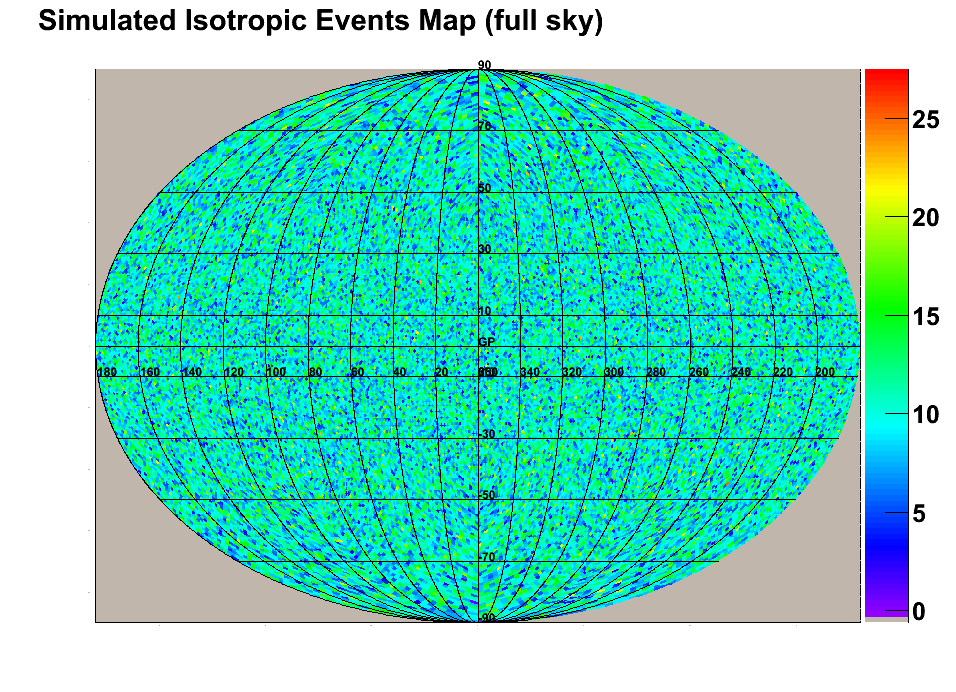} 	
 	\caption{Simulated background maps with ({\it left}) and without ({\it right}) acceptance, for the isotropic case.}
 	\label{fig:iso}
 \end{figure}

\begin{figure}[h!]
	\centering
 	\hspace{-1.7cm}
 	\includegraphics[scale=0.245]{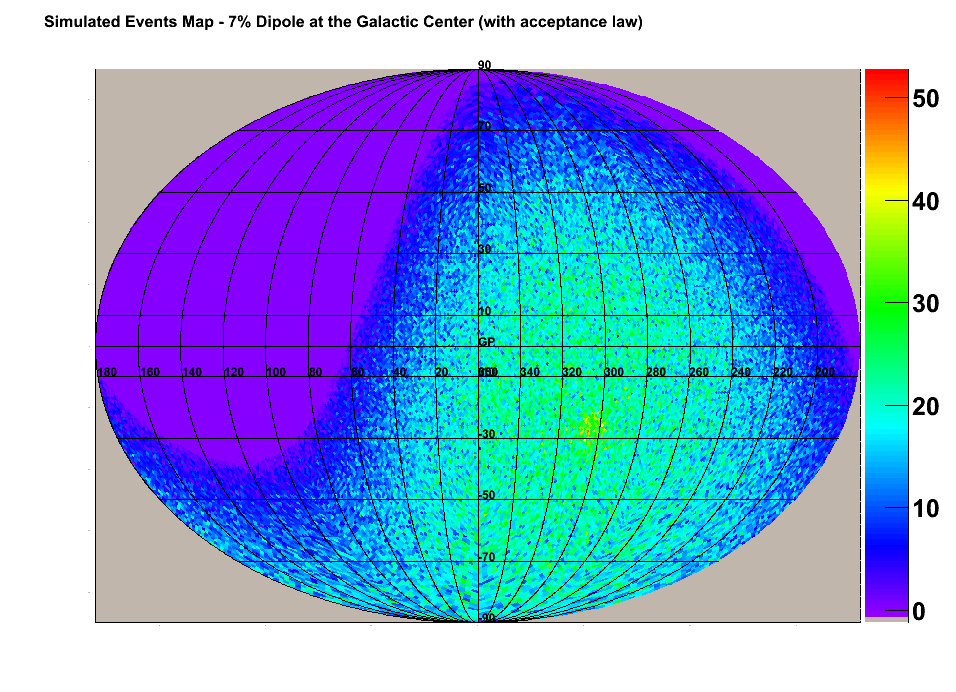} 
 	\includegraphics[scale=0.245]{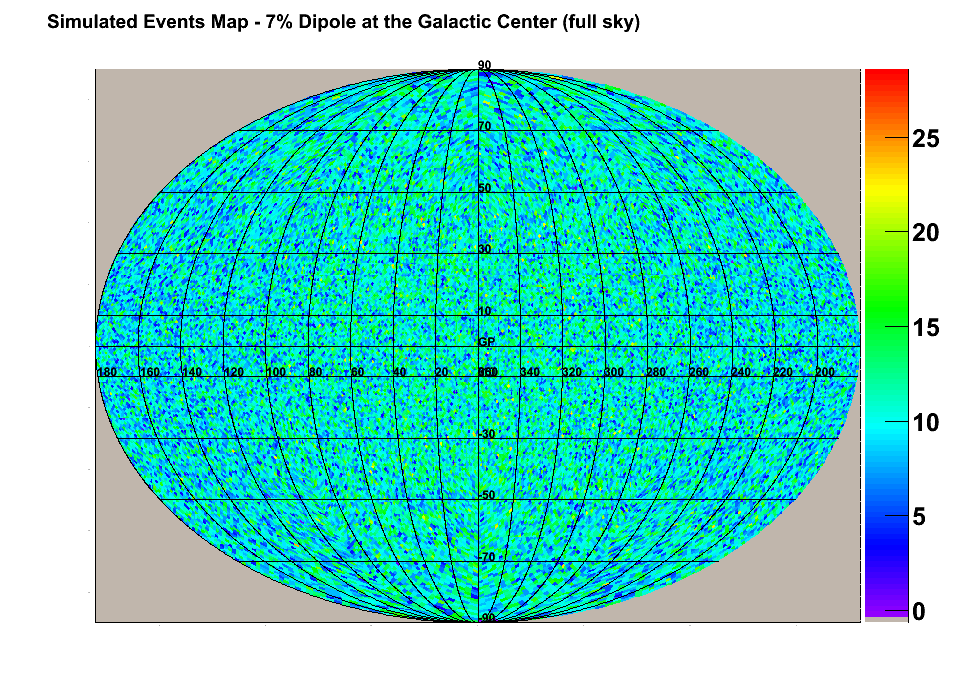} 	
 	\caption{Simulated background maps with ({\it left}) and without ({\it right}) acceptance, for the dipole 1 case.}
 	\label{fig:dipole1}
 \end{figure}

\begin{figure}[h!]
	\centering
 	\hspace{-1.7cm}
 	\includegraphics[scale=0.245]{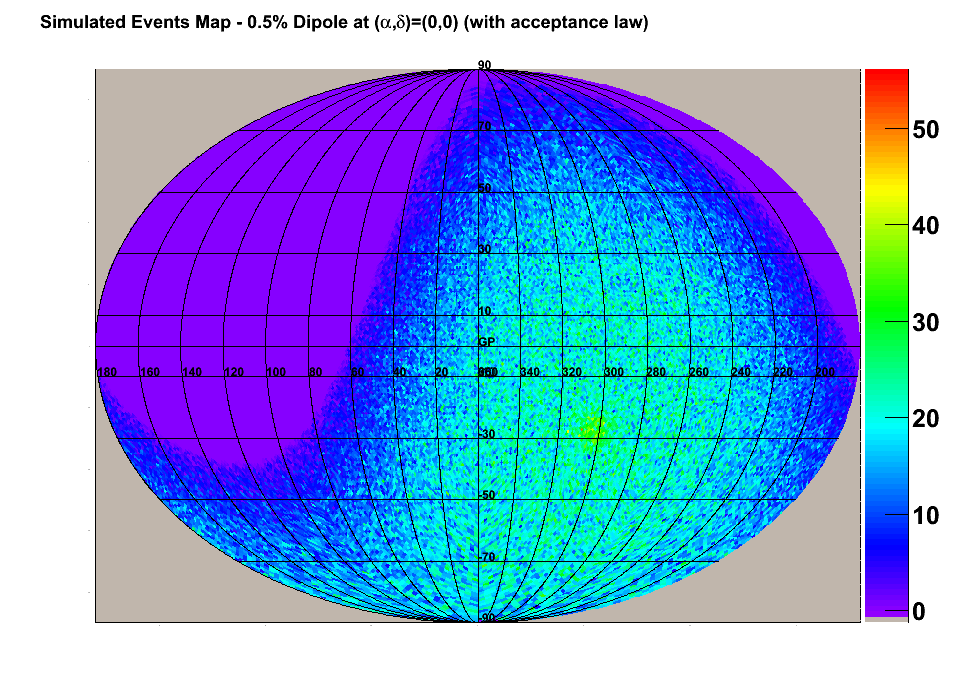} 
 	\includegraphics[scale=0.245]{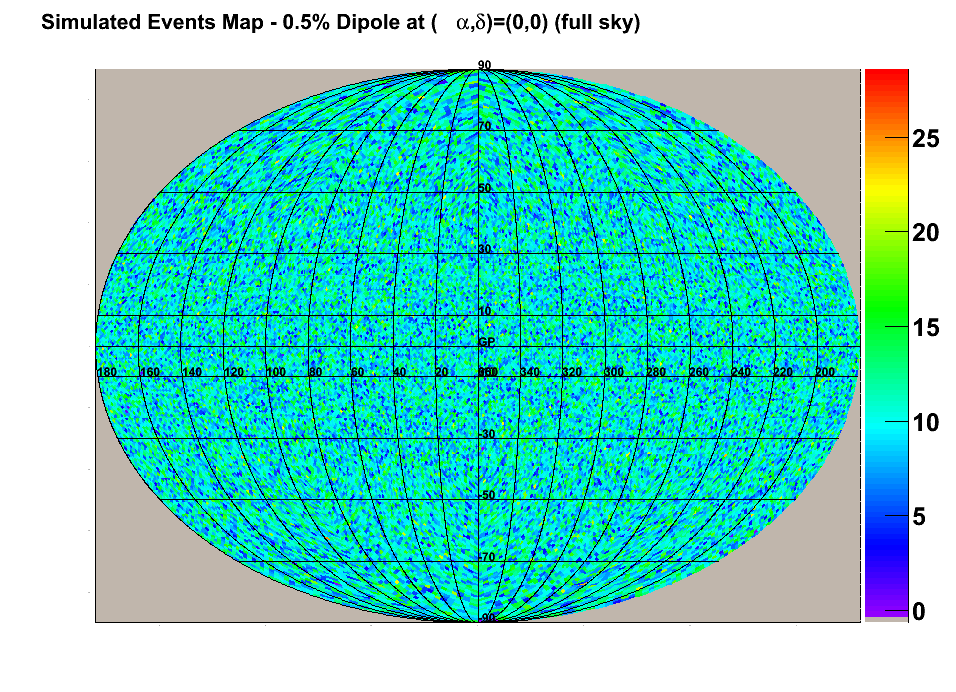} 	
 	\caption{Simulated background maps with ({\it left}) and without ({\it right}) acceptance, for the dipole 2 case.}
 	\label{fig:dipole2}
 \end{figure}

\begin{figure}[h!]
	\centering
 	\hspace{-1.7cm}
 	\includegraphics[scale=0.245]{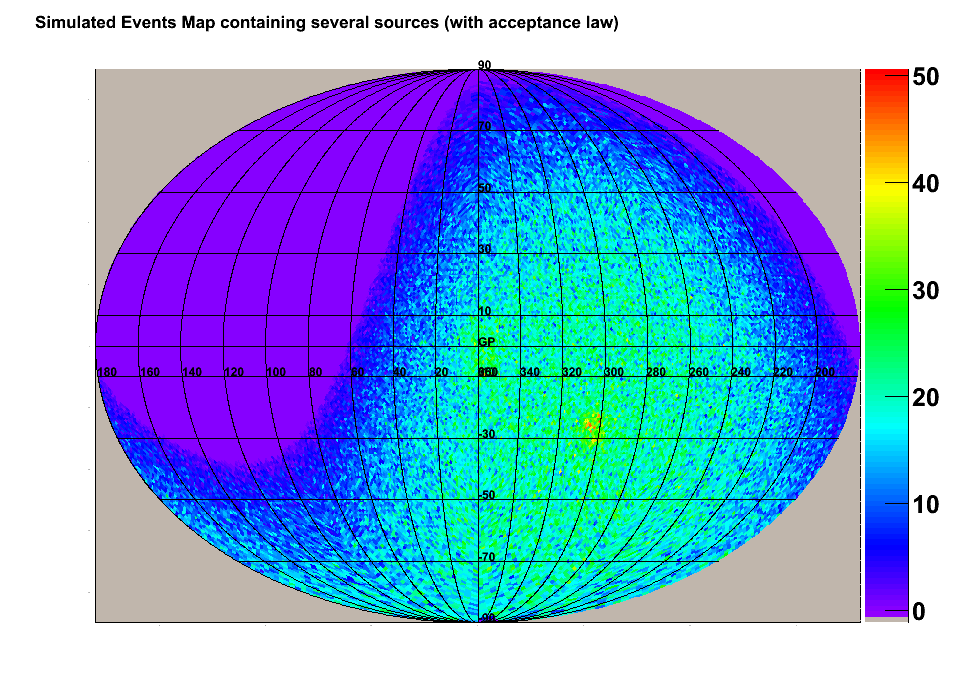} 
 	\includegraphics[scale=0.245]{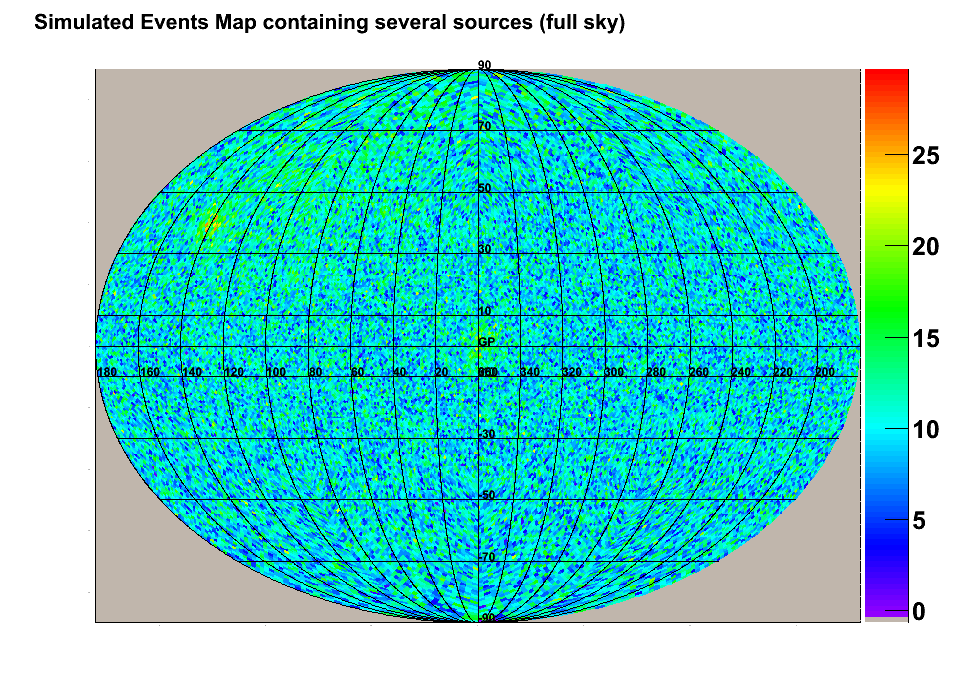} 	
 	\caption{Simulated background maps with ({\it left}) and without ({\it right}) acceptance, for the sources case.}
 	\label{fig:sources}
 \end{figure}

  Four anisotropy patterns were simulated as the background maps, with and without the acceptance of the detector (figures \ref{fig:iso} to \ref{fig:sources}). In all the cases these $500,000$ events were simulated. The simulated anisotropy patterns were:
 \begin{itemize}
 	\item {\bf isotropic:} isotropic distribution of events;
 	\item {\bf dipole 1:} a dipole with excess in the galactic center $(l,b)$ = $(0^o,0^o)$, with amplitude $7\%$ with respect to the background;
 	\item  {\bf dipole 2:} a dipole with excess in the direction $(l,b)$ = $(266.5^o,-29^o)$, with amplitude $0.5\%$ with respect to the background;
 	\item {\bf sources:} several sources with different angular scales $\sigma$ and amplitudes $A$ in the directions $(l,b)$: $(0^o,0^o)$ [$\sigma=7.0^o$, $A=100\%$], $(320^o,90^o)$ [$\sigma=1.5^o$, $A=5\%$],  $(320^o,-40^o)$ [$\sigma=0.5^o$, $A=1\%$],  $(220^o,10^o)$ [$\sigma=3.0^o$, $A=5\%$],  $(100^o,-70^o)$ [$\sigma=2^o$, $A=10\%$],  $(240^o,50^o)$ [$\sigma=20^o$, $A=5\%$],  $(350^o,-80^o)$ [$\sigma=6.0^o$, $A=0.5\%$],  $(100^o,50^o)$ [$\sigma=30^o$, $A=50\%$],  $(140^o,-40^o)$ [$\sigma=4.0^o$, $A=200\%$] and  $(60^o,50^o)$ [$\sigma=3.0^o$, $A=2\%$].
 \end{itemize}
 
 \par In order to verify the power of identification of the wavelets, we calculated the amplification ($\lambda$) of the signal-to-noise ratio, which is:
 \begin{equation}
 	\lambda = \frac{w_f/\sigma_f}{w_0/\sigma_0}, 
 	\label{eq:amp}
 \end{equation}
 where $w_0$ is the value of the central pixel associated to the source  in the non-filtered source map, $w_f$ is the value of the same pixel in the filtered source map, $\sigma_0$ is the root mean square (RMS) of the non-filtered background map and $\sigma_f$ is the RMS of the filtered background map.
 
 \par According to Gonz{\'a}lez-Nuevo {\it et al} \cite{mexicanhat}, the maximum amplification of the signal-to-noise ratio for a source with dispersion $\gamma_0$ embedded in a white noise\footnote{White noise has the property of being homogeneous and isotropic, with a uniform power spectrum.} background is obtained by convolving the maps with a gaussian kernel with dispersion $\gamma=\gamma_0$. This shall be used as a reference to check the consistency of our results.
 
\section{Results and Discussion}
 
 \par
\begin{figure}[h!]
	\centering
 	\hspace{-1.8cm}
 	\includegraphics[scale=0.252]{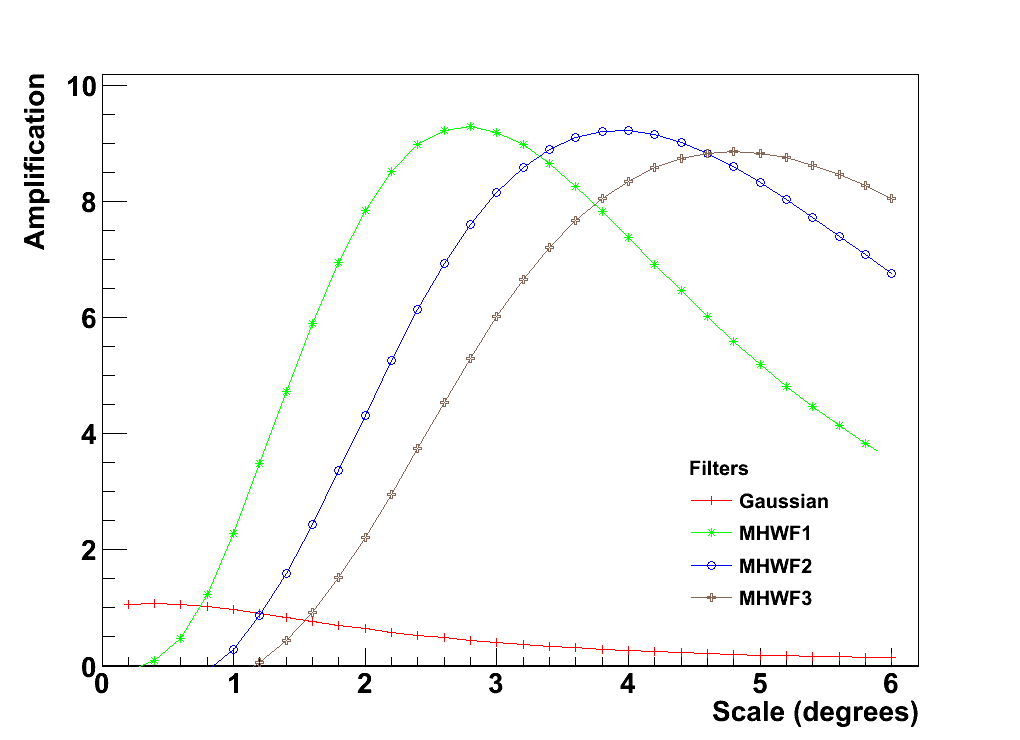} \hspace{-0.7cm}
 	\includegraphics[scale=0.252]{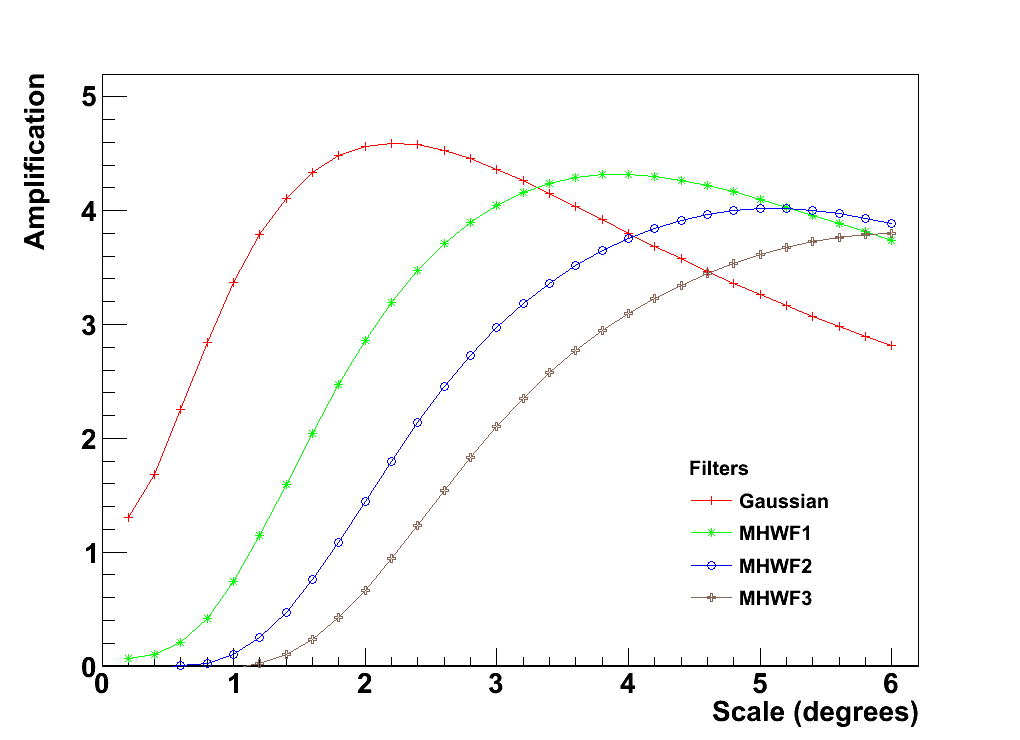} \hspace{-0.7cm} 	
 	\caption{Amplification for the isotropic case with ({\it left}) and without ({\it right}) acceptance.}
 	\label{fig:amp-iso}
    \end{figure}

\begin{figure}[h!]
	\centering
 	\hspace{-1.8cm}
 	\includegraphics[scale=0.252]{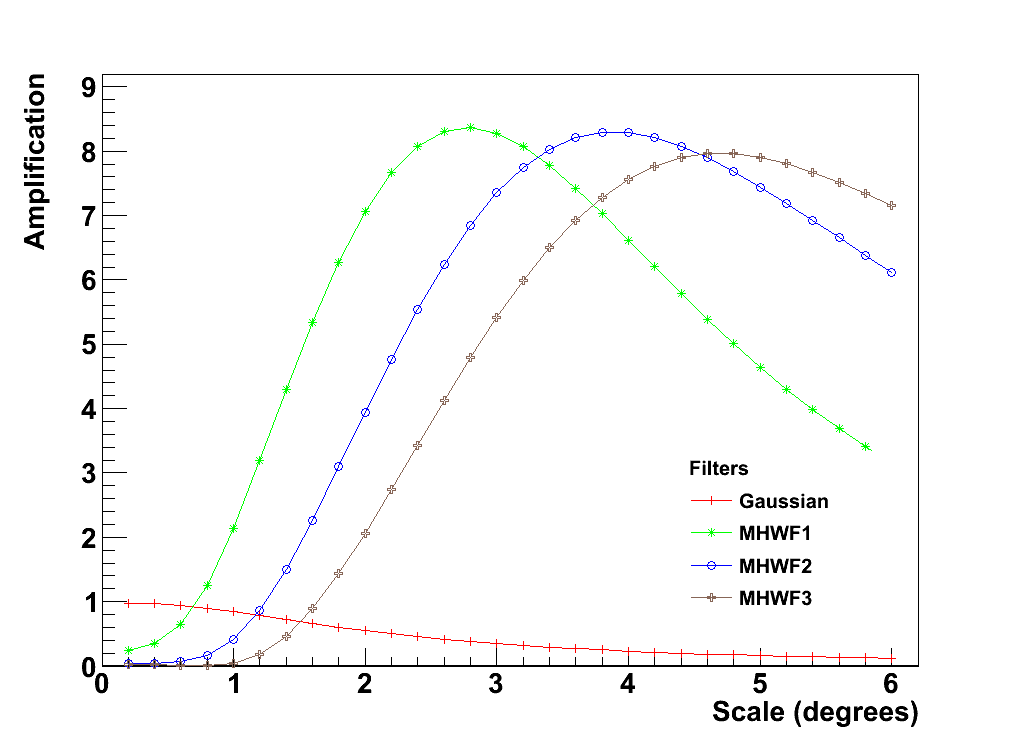} \hspace{-0.7cm}
 	\includegraphics[scale=0.252]{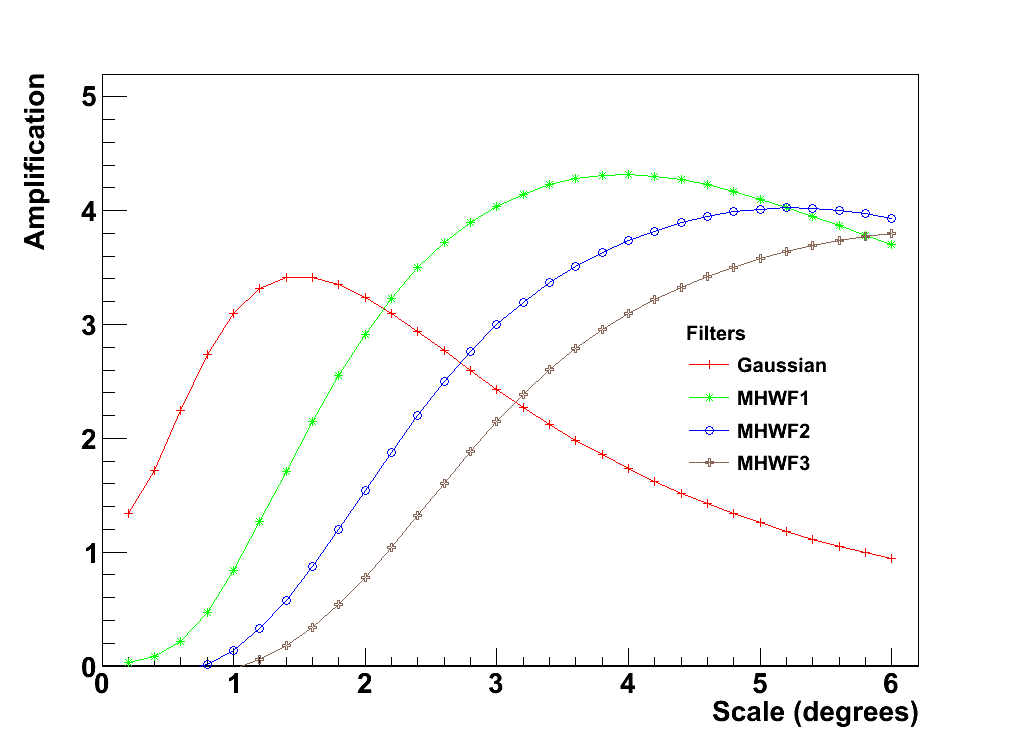} \hspace{-0.7cm} 	
 	\caption{Amplification for the dipole 1 case with ({\it left}) and without ({\it right}) acceptance.}
 	\label{fig:amp-dipole1}
    \end{figure}

\begin{figure}[h!]
	\centering
 	\hspace{-1.8cm}
 	\includegraphics[scale=0.252]{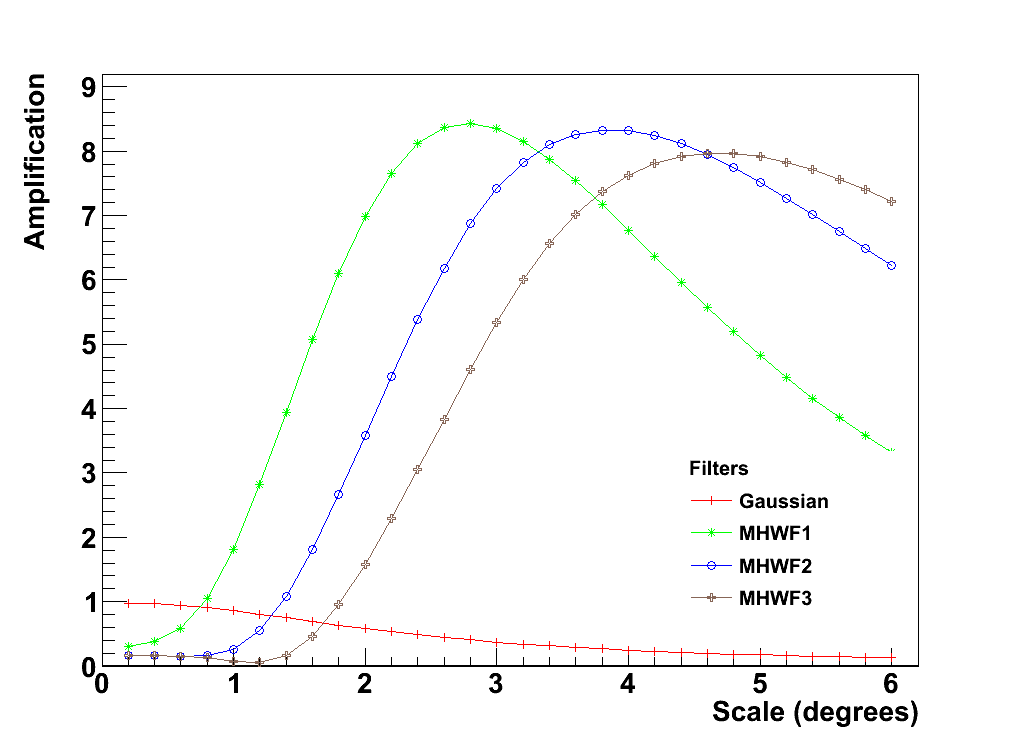} \hspace{-0.7cm}
 	\includegraphics[scale=0.252]{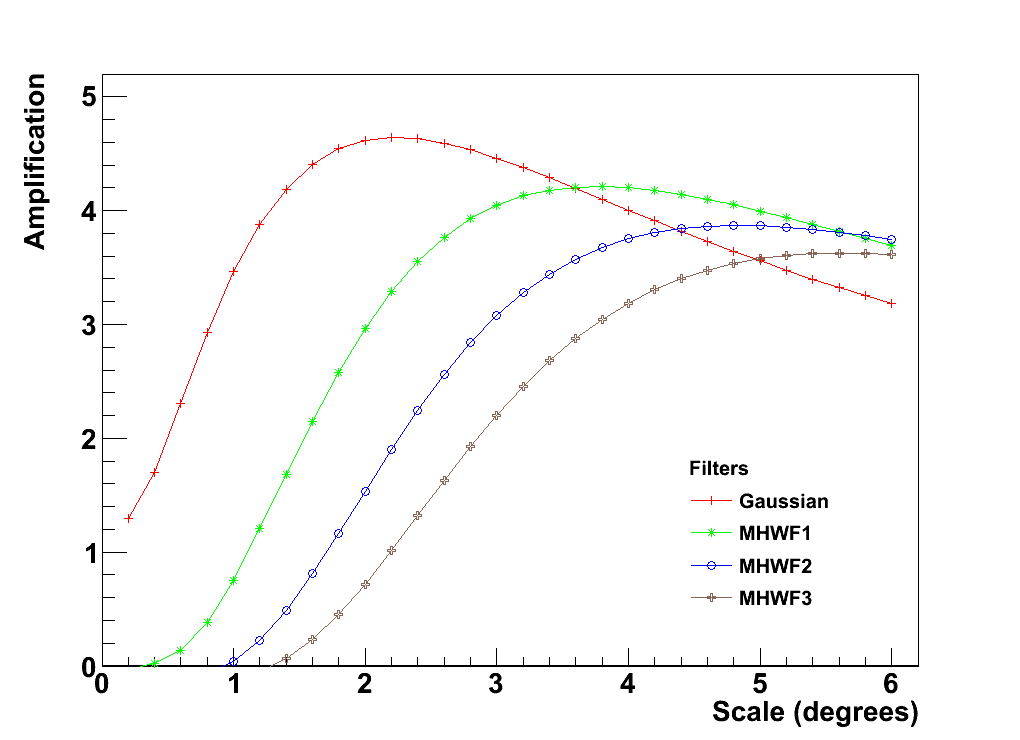} \hspace{-0.7cm} 	
 	\caption{Amplification for the dipole 2 case with ({\it left}) and without ({\it right}) acceptance.}
 	\label{fig:amp-dipole2}
    \end{figure}

\begin{figure}[h!]
	\centering
 	\hspace{-1.8cm}
 	\includegraphics[scale=0.252]{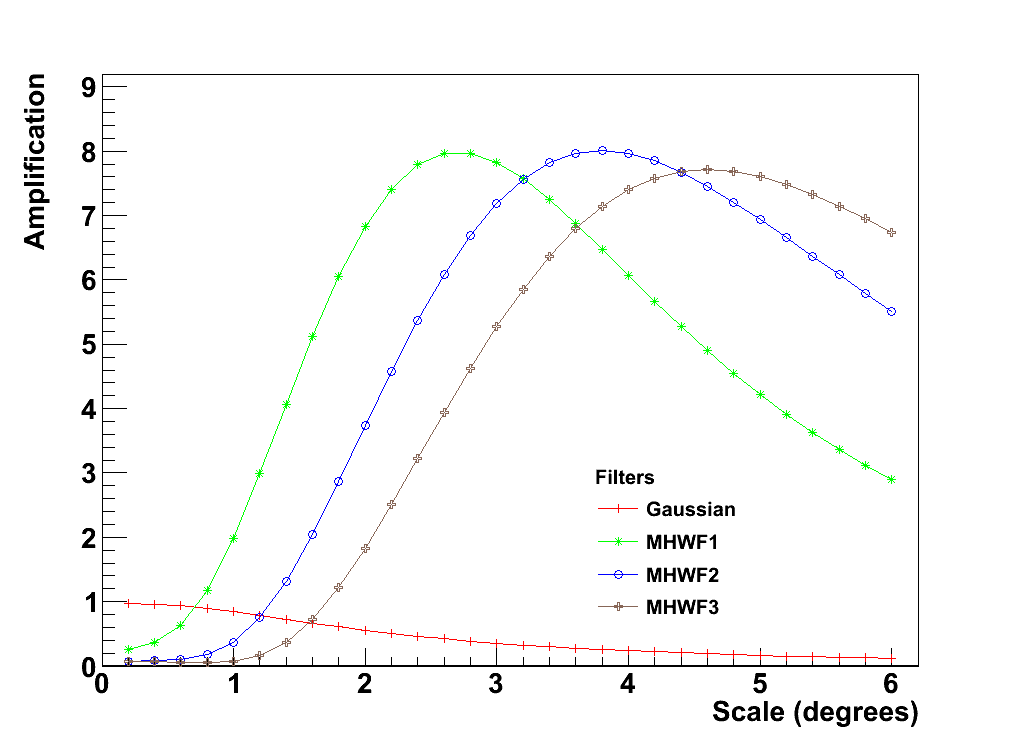} \hspace{-0.7cm}
 	\includegraphics[scale=0.252]{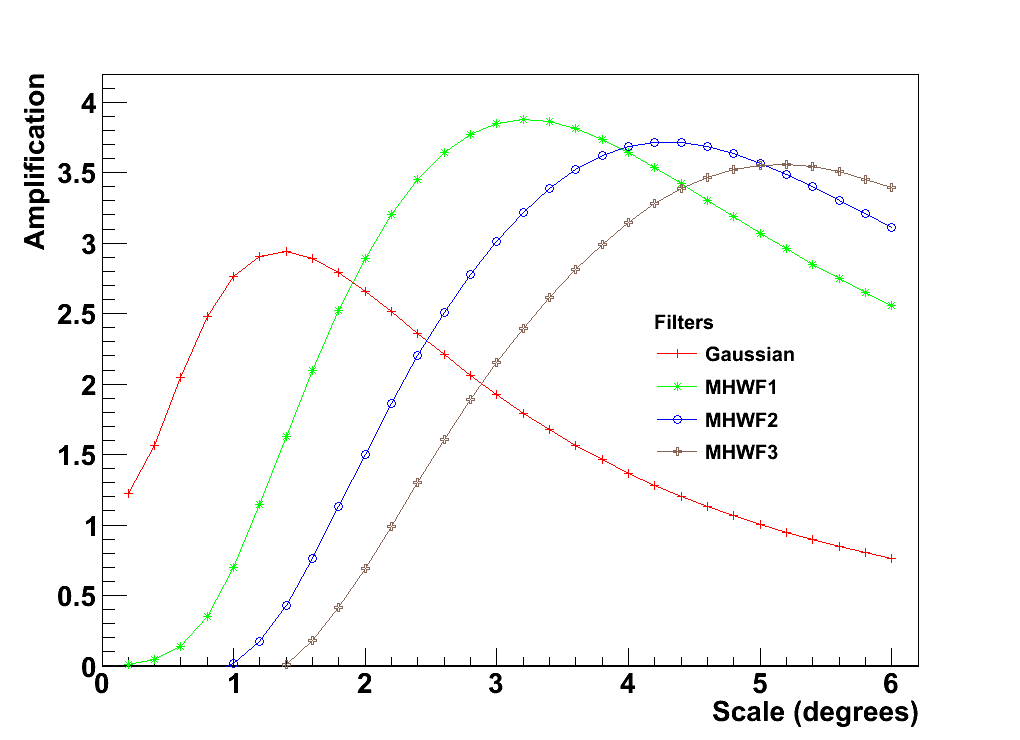} \hspace{-0.7cm} 	
 	\caption{Amplification for the simulated sources case with ({\it left}) and without ({\it right}) acceptance.}
 	\label{fig:amp-sources}
    \end{figure}

 Using equation \ref{eq:amp} we have calculated the amplifications in all the simulated cases, with scales spanning from $0.2^o$ up to $6.0^o$. We present here only the graphs of amplification in which the source has an amplitude of $1\%$ with respect to the background, with $5,000$ events in its direction. The results are shown in figures \ref{fig:amp-iso} to \ref{fig:amp-sources}.

\par We are interested in obtaining amplifications greater than one, indicating that the signal-to-noise ratio from the filtered map is greater than the one from the non-filtered map. For the first three orders of the MHWF, we have found amplifications greater than one for some scales. The same did not occur when we convolved the maps with gaussian filters taking into account the acceptance of the detector. In this case, the amplification for gaussian filters were less than one and in the best case, $10\%$ greater than one. However, if we reject the acceptance of the detector, the gaussian filter provides amplifications greater than one for some angular scales in all the cases. In figure \ref{fig:amp-iso} (right) we expect that the maximum amplification would be obtained for a gaussian filter with dispersion equal to the one from the simulated source, since this is the scenario of a source embedded in a white noise background. This can be clearly seen in this figure. In figure \ref{fig:amp-dipole1} (right) we have a dipole  with an excess of events $7\%$ greater than the background in the direction of the galactic center, without acceptance. Due to the effect of the imposed dipolar anisotropy pattern, the gaussian filter does not provide the maximum amplification of the signal-to-noise ratio. In the case of the second dipole without acceptance, shown in the right side of figure \ref{fig:amp-dipole2}, the gaussian filter provides the maximum amplification, since the simulated dipolar pattern is very faint ($0.1\%$).  In the case with sources and without acceptance, shown in figure \ref{fig:amp-sources} (right), the gaussian filter provides an amplification greater than one, but still less than the ones provided by the MHWF.

\par At first glance, the conclusion that the MHWF has a greater power of discrimination for sources when the background is not white noise seems reasonable. However, let us notice that the simulated source has a dispersion of $2.0^o$, and the peaks of amplifications are shifted to the right with respect to this value, in the case of the MHWF. So, the scale of the filter to be used and the scale of the source are not the same. Let $\gamma$ be the scale of the filter, corresponding to the dispersion of the gaussian of the used filter and $\gamma_0$ the dispersion of the simulated source, which in our case is $2^o$. We can define $\Delta \gamma = | \gamma - \gamma_{0} |$ as being the distance between the scale of maximum amplification and the dispersion of the simulated source. This is shown in tables \ref{tab:maxamp} and \ref{tab:maxampfs} .

\begin{center}
    \begin{table}
	\centering
	\caption{$\Delta \gamma$ (measured in degrees) for different anisotropies in the case of $N_{src}$ events from the source and an acceptance law for the detector.}
	\label{tab:maxamp}
	\begin{tabular}{cccccc} 
		\hline
		{\bf $N_{src}$} & {\bf Anisotropy} & {\bf Gauss} & {\bf MHWF1} & {\bf MHWF2} & {\bf MHWF3} \\
		\hline
		100,000   & isotropic & 1.8 & 0.8 & 2.0 & 2.8 \\ 
		100,000   & dipole 1  & 1.8 & 0.8 & 2.0 & 2.8 \\   
		100,000   & dipole 2  & 1.8 & 0.8 & 1.8 & 2.6 \\  
		100,000   & sources  & 1.8 & 0.8 & 1.8 & 2.6 \\    
		5,000   & isotropic & 1.6 & 0.8 & 2.0 & 2.8 \\ 
		5,000   & dipole 1  & 1.8 & 0.8 & 1.8 & 2.6 \\   
		5,000   & dipole 2  & 1.8 & 0.8 & 1.8 & 2.6 \\  
		5,000   & sources  & 1.8 & 0.8 & 1.8 & 2.6 \\    
		50   & isotropic & 1.8 & 0.8 & 2.0 & 2.8 \\ 
		50   & dipole 1  & 1.0 & 0.8 & 1.8 & 2.6 \\   
		50   & dipole 2  & 0.2 & 1.0 & 2.2 & 3.2 \\  
		50   & sources  & 1.4 & 0.6 & 1.8 & 2.6 \\    
	\end{tabular}
    \end{table}
\end{center}	

\begin{center}
    \begin{table}
	\centering
	\caption{$\Delta \gamma$ (measures in degrees) for different anisotropies in the case of $N_{src}$ events from the source and no acceptance law.}
	\label{tab:maxampfs}
	\begin{tabular}{cccccc} 
		\hline
		{\bf $N_{src}$} & {\bf Anisotropy} & {\bf Gauss} & {\bf MHWF1} & {\bf MHWF2} & {\bf MHWF3} \\
		\hline
		100,000   & isotropic & 0.0 & 1.8 & 3.0 & 4.0 \\ 
		100,000   & dipole 1  & 0.4 & 2.0 & 3.2 & 4.0 \\   
		100,000   & dipole 2  & 0.2 & 1.8 & 2.8 & 3.6 \\  
		100,000   & sources  & 0.6 & 1.2 & 2.2 & 3.2 \\    
		5,000   & isotropic & 0.2 & 2.0 & 3.0 & 4.0 \\ 
		5,000   & dipole 1  & 0.4 & 2.0 & 3.2 & 4.0 \\   
		5,000   & dipole 2  & 0.2 & 1.8 & 2.8 & 3.6 \\  
		5,000   & sources  & 0.6 & 1.2 & 2.2 & 3.2 \\    
		50   & isotropic & 0.4 & 1.6 & 2.8 & 3.8 \\ 
		50   & dipole 1  & 0.6 & 1.8 & 3.2 & 4.0 \\   
		50   & dipole 2  & 0.2 & 1.2 & 2.0 & 2.8 \\  
		50   & sources  & 1.0 & 0.8 & 1.8 & 2.6 \\    
	\end{tabular}
    \end{table}
\end{center}		

\par We notice in table $\ref{tab:maxamp}$  that $\Delta \gamma$ for different anisotropies is approximately constant for all filters, despite the anisotropy pattern of the background. However, for 50 events coming from the direction of the source, we notice that $\Delta \gamma$ fluctuates up to $1.6^o$ in the case of the gaussian filter and at most $0.6^o$  for the MHWF filters. The same is not true for the full sky simulations, presented in table \ref{tab:maxampfs}. In this case,  $\Delta \gamma$ fluctuates up to $1.4^o$ for the MHWF filters and $0.8^o$ for the gaussian kernel.

\section{Conclusions}

\par In this work we have considered a $sin\theta cos\theta$ law for the zenith angle distribution for a virtual detector located at the same place as the Pierre Auger Observatory and an ideal detector observing the whole sky, with uniform acceptance. We have simulated a source with different number of events coming from it, embedded in several backgrounds. We have calculated the amplifications of the signal-to-noise ratio in each case and verified that in the case of a source with dispersion $\gamma_0$ embedded in a white noise background, the maximum amplification is achieved with a gaussian filter with dispersion $\gamma=\gamma_0$, as predicted. In the case of a faint dipole with no acceptance, the gaussian provided a greater amplification of the signal-to-noise ratio. However, if the background does not have a constant power spectrum, as in the case of the other anisotropy patterns, the Mexican Hat Wavelet Family filters provided, in general, a greater amplification of the signal-to-noise ratio, allowing one to identify point sources in maps containing arrival directions of cosmic rays. 

\par It is interesting to notice that the existence of an acceptance for the detector affects the power of discrimination of the gaussian filter. The amplification, in this case, is always below $1.1$, whereas in the case of the MHWF the amplification of the signal-to-noise ratio can be more than ten. 


\par The acceptance of the experiment affects the power spectrum of the background in celestial maps. If it is not white noise, the maximum amplification of the signal-to-noise ratio is achieved by the MHWF kernel. However,  even though we can amplify the signal-to-noise ratio using MHWF, we lose directional information about the associated cosmic ray events. If we consider a gaussian filter and a white noise background, the optimal gaussian filter would have exactly the same dispersion as the source and the uncertainty on the position would arise  from the resolution of the detector, rather than the analysis method used. By using MHWF, the greater the order of the wavelet, the greater the difference between the sizes of the source and the wavelet. Therefore, despite the gain on the power of discrimination of the source, there is a loss of resolution.

\par Since the acceptance introduces a non-white noise background, for analysis involving the whole sky, such as blindsearch\footnote{The procedure of blindsearch is to search for astrophysical objects on a given window around any observed excess.}, the MHWF amplifies the signal-to-noise ratio more than the gaussian. However, for small scale analysis, smoothing the maps with a gaussian kernel and reducing the area of scan could provide a greater amplification, specially if the background within the window of scan has a uniform power spectrum.


\par A limitation to this technique is the projection of the celestial sphere to the plane. This approximation is good enough for small scales, when we can use the approximation $sin\theta \approx \theta$. For $\theta > 7^o$, this approximation introduce bias and this method would have to be adapted to work on a spherical manifold. Wavelets on the sphere have been used on CMB studies\cite{wiaux,mcewen1,mcewen2,mcewen3} and presented good results. The next step of this work is a similar analysis using wavelets on the sphere, which allow us to search not only point sources, but also large scale structures in maps containing arrival directions of cosmic rays.

\section{Acknowledgements}

We are grateful for the financial support of FAPESP (Funda{\c{c}}{\~a}o de Amparo \`a Pesquisa do Estado de S{\~a}o Paulo) and CNPq (Conselho Nacional de Pesquisa e Desenvolvimento). Also, we would like to thank the developers of the {\it Coverage \& Anisotropy Toolkit}, which allowed us to perform fast simulations of cosmic ray events.


\end{document}